# Thermal Probing of Energy Dissipation in Current-Carrying Carbon Nanotubes

Li Shi,[1*] Jianhua Zhou,[1] Philip Kim,[2] Adrian Bachtold,[3] Arun Majumdar,[4,5] Paul L. McEuen[6]

[1]Department of Mechanical Engineering and Center for Nano and Molecular Science and Technology, The University of Texas at Austin, Austin, TX 78712, USA

[2]Department of Physics, Columbia University, New York, NY 10027, USA

[3]Centre d'Investigacions en Nanociencia i Nanotecnologia (Consejo Superior de Investigaciones Científicas–Institut Català de Nanotecnologia), Campus de la Universitat Autònoma de Barcelona (UAB), E-08193 Bellaterra, Spain

[4]Department of Mechanical Engineering, University of California, Berkeley, CA 94720, USA;

[5]Materials Science Division, Lawrence Berkeley National Laboratory, Berkeley, CA 94720, USA

[6]Laboratory of Atomic & Solid State Physics, Cornell University, Ithaca, NY 14853 USA

*Corresponding author. E-mail:lishi@mail.utexas.edu

ABSTRACT

The temperature distributions in current-carrying carbon nanotubes have been measured with a scanning thermal microscope. The obtained temperature profiles reveal diffusive and dissipative electron transport in multi-walled nanotubes and in single-walled nanotubes when the voltage bias was higher than the 0.1–0.2 eV optical phonon energy. Over ninety percent of the Joule heat in a multi-walled nanotube was found to be conducted along the nanotube to the two metal contacts. In comparison, about eighty percent of the Joule heat was transferred directly across the nanotube-substrate interface for single-walled nanotubes. The average temperature rise in the nanotubes is determined to be in the range of 5 to 42 K per micro watt Joule heat dissipation in the nanotubes.

I. INTRODUCTION

The intriguing one-dimensional (1-d) nature of electrons in carbon nanotubes (NTs) have attracted considerable amount of theoretical and experimental work. Transport studies suggest that the electron mean free path can be remarkably long in single-walled (SW) NTs at low electrical bias. [1-9] On the other hand, transport measurements of multi-walled (MW) NTs [10] and SWNTs [4,11-16] under high bias voltage have suggested that the energy dissipation is largely due to the increased coupling of these energetic electrons to optical and zone boundary phonons, and hence the increase of lattice temperature is expected. The resulted temperature distribution could play an important role in electron transport in NTs and their possible applications for nanoelectronics and interconnects.

The issue of energy dissipation in NTs has received increasing attention in recent years and has been tackled by means of several experimental approaches. For instance, the electric current-voltage (*I-V*) characteristics of suspended SWNTs were measured and fitted to a coupled electron-phonon transport model.[17] The lattice temperature rise and the thermal conductivity of the NT were extracted by adjusting several fitting parameters. In two other reports, [18,19] self Joule heated NTs were employed as both heaters and electrical resistance thermometers to obtain the lattice thermal conductivity. The inconvenience of such approaches is that the results obtained on the thermal properties depend on the model employed as well as on several parameters that are difficult to characterize, such as the coupling between the optical and acoustic phonons and the contact electrical and thermal resistances at the nanotube-electrode interfaces. A more direct method has recently been reported. This method is based on micro-Raman spectroscopy that measures the optical phonon temperature of suspended SWNTs under high bias voltage.[20]

A different experiment in a transmission electron microscope (TEM) has been used to study the temperature profile along a self Joule heated MWNT.[21] In this work, the melting and evaporation of nanocrystal thermometers were imaged to map the temperature distribution of a suspended silicon nitride membrane supporting the MWNT.[21] A finite element model based on a 1d heat diffusion equation was applied to extract the thermal conductivity and temperature rise of the MWNT. In the model the thermal interface resistance between the MWNT and the supporting membrane was neglected.

In this paper, we present measurements of the local lattice temperature of voltage biased NTs using a microscopic thermocouple junction at the end of an atomic force microscope (AFM) tip. These measurements allow us to study how the Joule heat is transferred out of the NT. Most of the heat is found to exit MWNTs through the electrodes. In contrast, the heat dissipated in SWNTs is primarily transferred directly through the substrate. These measurements also enable the evaluation of the lattice temperature rise of voltage-biased NTs as well as the thermal resistances at the tube-substrate interface and at the tube-electrode interface.

## II. MEASUREMENT METHOD AND RESULTS

Scanning thermal microscope (SThM) tips have been fabricated and characterized as described in detail elsewhere.[22,23] As shown in Fig. 1a, the micro-patterned Pt and Cr lines form a junction at the apex of a $SiO_2$ tip that has a tip radius of about 30 nm. It is known from the previous study[23,24] that the spatial resolution of the thermal probe can be as small as about 50 nm. Compared to heat conduction through the solid-solid contact between the tip and the sample and a liquid meniscus around the tip, radiation heat transfer was found to be negligible.[25] On the other hand, heat transfer via the air gap between the thermocouple tip and the sample makes a noticeable contribution to the measured tip temperature. When the line width of a Joule heated metal line sample was reduced from 5.8 μm to about 350 nm, the air contribution was found to decrease and direct heat conduction through the solid-solid contact and the liquid meniscus became the dominant tip-sample heat transfer mechanism.[24] Although the much narrower line width or diameter of the Joule heated NT sample in this study is expected to decrease the air contribution further, the measured tip temperature is still given by the convolution of the NT temperature and the surface temperature of the nearby substrate due to heat transfer through the solid-solid contact, liquid meniscus, and air at the contact point. In addition, to protect the SWNTs during contact mode scanning of the SThM probe, a ~5 nm polystyrene film was spun on the SWNT samples. The electrically insulating polystyrene film together with the thin oxide layer formed on the outer Cr layer of the thermal probe prevent direct electronic coupling between the probe and the sample. Hence, the measured thermal probe temperature rise is caused by the lattice temperature rises of the NTs and nearby substrate. While the parasitic heat transfer between the substrate and the tip via different mechanisms deteriorate the spatial resolution to be larger than the NT diameter, the measured axial temperature profile along a several micron long NT is expected to be approximately proportional to that in the NT with a spatial resolution on the 100 nm scale, provided that the thermal coupling between the NT and the substrate remains relatively constant along the NT.

Shown in Fig. 1b is the topographic image of a 10 nm diameter MWNT sample contacted by two Au/Cr contacts patterned on top of the nanotube using electron beam lithography and a metal lift-off technique. Figure 1c shows the obtained thermal image at 0.73 V DC bias and 20.1 μA current. As revealed in Fig. 1d, the temperature profile along the MWNT clearly shows that the highest temperature occurs in the middle of the MWNT, as expected for diffusive heating of a metal wire. Similar temperature profiles have been measured for other MWNTs. [26,27] On the other hand, the source-drain current ($I_{sd}$) versus voltage ($V_{sd}$) curve in Fig. 1e reveals that the conductance increased when the bias was increased to be above about 0.5 V. Bourlon *et al.* has reviewed this behavior and attributed it to an increase of the number of current-carrying shells and/or to electrons subjected to Coulomb interaction that tunnel across the MWNT-electrode interface. [28]

We now turn our attention to SWNTs. Figure 2(a-b) shows the AFM height image and thermal image of a 1.8 nm diameter SWNT at a DC bias of 1.7 V and current 15.1 µA. The DC thermal image of a SWNT is generally noisier than that of a MWNT because the smaller diameter of the SWNT leads to less thermal coupling between the NT and the thermal probe. Nevertheless, diffusive electrical heating can still be observed in this and other SWNT samples at a DC bias voltage of about 0.4 V or higher corresponding to an electrical heating rate larger than 1 µW per µm NT length. At a lower heating rate, the signal to noise ratio was too low to be detected by the thermal probe.

In order to improve the thermal signal, we coupled a sinusoidal current to the DC current and used a lock-in amplifier to measure the first harmonic oscillation in the SThM signal. The AC voltage amplitude was kept to be smaller than the DC voltage so that electrons drift always along the same direction. This AC detection approach greatly improves the signal to noise ratio, as shown in Fig. 2c for an applied bias of $(0.767 + 0.733 \sin 2\pi f t)$ V and $f = 205$ Hz. The corresponding temperature profile in Fig. 2d along the SWNT clearly shows that the highest temperature occurred near the center of the nanotube.

Figure 3(b-d) show that the AC thermal signal could be observed for voltage bias larger than 0.22 V applied to a 1.2 nm diameter, 3.3 μm long metallic SWNT with a low-bias resistance of 41 kΩ. For another 2 nm diameter, 2.1 μm long SWNT sample with a low-bias resistance of 22 kΩ, diffusive heating was observed in the AC thermal image when the voltage bias was higher than 0.12 V (not shown here). When the voltage bias was further reduced, the thermal probe was not able to detect heating above the noise level. Here, the onset of the observed diffusive heating coincides with the 0.1–0.2 V minimum bias that is needed to generate hot electrons of sufficient energy for optical phonon emission, which tends to saturate the electron current at high voltage bias as described in previous works[11] and revealed in Fig. 3a. Further improvement in the sensitivity of the SThM is needed to verify whether there exists a cross-over from non-dissipative to dissipative transport at 0.1–0.2 V bias.

### III. ANALYSIS AND DISCUSSION

While diffusive heating was observed in both MWNTs and SWNTs at high biases, the temperature profiles along the MWNTs and the SWNTs are distinctly different. Although the measured DC temperature profile of Fig. 1d is noisier than the measured AC temperature profile of Fig. 2d., it is apparent that the temperature gradient near the center of the MWNT in Fig. 1d is much higher than that for the SWNT in Fig. 2d For the SWNT, the temperature remains approximately flat over two third of the nanotube length near the center and starts to decrease toward the two contacts at a distance ($L_c$) of about 0.5 μm from each contact. For the MWNT in Fig. 1d and several longer MWNTs,[26,27] such a flat top in the temperature profile was not observed in the measured temperature distribution, and thus $L_c$ could be longer than 2 μm.

To understand this difference, we employ the following heat diffusion equation to investigate how the Joule heat dissipated in the nanotube is evacuated through the electrodes and the substrate

$$kA\frac{\partial^2 T}{\partial x^2} + \frac{Q}{L} - g(T-T_0) = \rho CA\frac{\partial T}{\partial t}, \tag{1}$$

where $T$, $k$, $A$, $L$, $Q$, $\rho$, and $C$ are the local lattice temperature, thermal conductivity, cross section area, length of the NT between the two metal electrodes, total electrical heating rate, density, and specific heat of the nanotube, respectively. The parameter $g$ is the nanotube-substrate thermal conductance per unit length, $T_0$ is the ambient temperature, and the axial distance $x$ is measured from the center of the tube, as illustrate in Fig. 4. The steady state solution of the normalized axial temperature distribution for the DC electrical heating case is given by

$$\theta(x) \equiv \frac{T(x)-T_0}{T_m-T_0} = \left(1-\frac{\cosh(mx)}{\cosh(mL/2)}\right)q^* + \frac{T_1+T_2-2T_0}{2(T_m-T_0)}\frac{\cosh(mx)}{\cosh(mL/2)} + \frac{T_2-T_1}{2(T_m-T_0)}\frac{\sinh(mx)}{\sinh(mL/2)},$$

$$(2)$$

where $T_m$ is the mean temperature along the nanotube, $T_1$ and $T_2$ are the nanotube temperatures at the two metal contacts, $q^* \equiv Q/Lg(T_m-T_0)$ is the dimensionless heating rate, and $m \equiv \sqrt{g/kA}$. Note that $1/m$ has the dimension of length and is expected to be close to $L_c$.[29] We notice that the frequency (~ 205 Hz) in our AC SThM measurements is low enough so that the AC measurement results can also be analyzed using the steady state solution from Eq. 2.[30]

The measured temperature profiles along the nanotube were fitted using Eq. 2. Shown in Fig. 2d, the $q^*$ and $m$ parameters obtained with least square fitting are $1.14 \pm 0.01$ and $3.0 \pm 0.1$ $\mu m^{-1}$, respectively. The rate of heat conduction ($Q_C$) along the nanotube to the two electrodes can be calculated from the temperature profile as

$$\frac{Q_C}{Q} = \frac{2\left(1 - \frac{\Delta T_1 + \Delta T_2}{2\Delta T_m q^*}\right)\tanh\left(\frac{mL}{2}\right)}{mL}, \qquad (3)$$

where $\Delta T_i = T_i - T_0$. For the four 1.2-2.0 nm diameter, 2.1-3.3 μm long SWNT samples measured in this work, the obtained $Q_C/Q$ ratio is between 12% and 23% and the remaining 77%-88% is dissipated directly across the SWNT-substrate contact to the substrate. For the four SWNTs, we did not observe a clear dependence of the $Q_C/Q$ ratio on the NT diameter and length. In comparison, the obtained $Q_C/Q$ ratio was found to be larger than 90% for the 10 nm diameter, 1.7 μm long MWNT sample in Fig. 1. We attribute the larger $Q_C/Q$ ratio of the larger diameter MWNT mainly to the smaller ratio between the tube-substrate contact area and the cross section area, which results in a smaller fraction of heat current through the tube-substrate interface.

The average temperature rise along the NT is obtained as $\Delta T_m = Q/kAm^2Lq^*$. The inset of Fig. 2d shows an approximately linear relationship between the obtained $k\Delta T_m$ and $Q$ for several AC and DC heating cases of the SWNT. For the case of $Q = 12.5$ μW, $\Delta T_m$ would be between 210 and 70 K if one assumes the reported measured thermal conductivity values between 1000 and 3000 W/m-K of individual SWNTs grown under similar conditions using thermal chemical vapor deposition (CVD).[17,31] Based on the dimensionless temperature profile in Fig. 2d, the maximum lattice temperature rise ($\Delta T_{max}$) near the center of the nanotube would be 240- 80 K and the temperature rise near the two electrodes ($\Delta T_c$) would be 90-30 K. For the MWNT in Fig. 1c, we obtain $\Delta T_m$ ≈540-190 K, $\Delta T_{max}$ ≈ 798-266 K, and $\Delta T_c$ ≈ 285-95 K at $Q$ = 14.7 µW using the measured thermal conductivity range of 1000-3000 W/m-K of an individual MWNT of ~10 nm diameter and grown by the arc discharge method used to produce the MWNTs in this study.[32-34] We note that lower thermal conductivity has been measured for an individual MWNT grown by CVD,[18] which could yield different crystal quality from MWNTs synthesized by the arc method. Moreover, a thermal conductivity lower than 1000 W/m-K would result in $\Delta T_{max}$ exceeding the breakdown temperature of about 900 K for a MWNT in air.[10,35] Since breakdown was not observed in the MWNT when the voltage was increased above the 0.73 V used in the thermal imaging experiment, we conclude that the thermal conductivity of the MWNT should not be lower than 1000 W/m-K. Despite the uncertainty in the NT thermal conductivity, these results show that the temperature rise of the NT at the electrodes is non-negligible, in contrast to what is often assumed.

The maximum lattice temperature rise that we found for the SWNT is about a factor of 2.7-8 lower than the recently reported ~650 K temperature of optical phonons probed by micro-Raman spectroscopy in the center of a suspended SWNT at a similar Joule heating rate.[20] This difference can be attributed to the elimination of heat transfer through the substrate for the suspended SWNT and non-equilibrium between the optical and acoustic phonon temperatures, the latter of which makes a large contribution to thermal signal detected by the thermal probe.

The overall thermal resistance between the NT and the ambient environment can be obtained as $R_{th} \equiv \Delta T_m / Q = 1 / m^2 k q^* A L$. The $R_{th}$ value based on $k \approx$ 1000-3000 W/m-K is in the range of 5.4 to 52 K/µW for four SWNTs and 13-39 K/µW for the MWNT in

Fig. 1c. $R_{th}$ is influenced by $k$, g, and the thermal interface resistance ($R_{i,e}$) between the NT and the electrodes.

Based on the extracted $m$ value, the nanotube-substrate conductance per unit length can be obtained as $g=m^2kA$. If one assumes $k \approx 1000\text{-}3000$ W/m-K, the obtained $g$ value would be in the range of 0.007 and 0.06 W/m-K for the four SWNT samples. The thermal conductance $g$ per unit NT length consists of contributions from the NT-SiO$_2$ interface resistance ($R_i$) per unit length, the spreading resistance ($R_S$) per unit length in the SiO$_2$ film and the Si substrate, i.e. $1/g = R_i + R_S$. We calculate that $R_S$ is smaller than 1.4 K m/W,[36] much smaller than the $1/g$ value in the range of 17-142 K m/W. Hence, $g$ is dominated by the interface resistance, i.e., $g \approx 1/R_i$. In addition, the obtained $g$ range is four to thirty times lower that the 0.2 W/m-K value that was determined using electrical breakdown of SWNTs at a much higher temperature for initiating the breakdown.[37,38] The lower $g$ value that we find may be attributed to the temperature dependence of the thermal interface resistance.[39] Indeed, the temperature in our case is lower than that during electrical breakdown and the nanotube-substrate interface resistance is proportional to the specific heat, which increases with temperature.

The interface thermal resistance between the NT and the electrodes is calculated as $R_{i,e} \equiv (\Delta T_1 + \Delta T_2)/Q_C = [(\Delta T_1 + \Delta T_2)/\Delta T_m]R_{th}(Q/Q_C)$. If $k \approx 1000\text{-}3000$ W/m-K, the $R_{i,e}$ values would range from 14 to 138 K/μW for four SWNTs and would be about 5-15 K/μW for the 10 nm diameter MWNT. The smaller $R_{i,e}$ for the MWNT is due to the larger circumference contact area between the MWNT and the electrode. In addition, the slightly asymmetric temperatures at the two contacts to the MWNT in Fig. 1d could be caused by a variation of $R_{i,e}$ at the two contacts or the presence of asymmetric Joule heating at the two contacts, which has not been taken into account in the above analysis.

## IV. CONCLUSIONS

This experiment provides detailed information of energy dissipation mechanisms in NTs. The measurement results show that most of the Joule heat in a SWNT is conducted directly to the substrate; whereas such direct heat transfer to the substrate could be substantially lower in a MWNT because of the smaller ratio between the nanotube-substrate contact area to cross section area. The measurement also reveals that the NT temperature rise near the electrodes was about half of the average temperature rise, which is determined to be in the range of 5 to 42 K per µW Joule heat dissipation in the nanotube. In addition, the NT-substrate interface thermal resistance was determined to be much larger than the spreading thermal resistance of the substrate. These contact and interface effects should be taken into account in modeling coupled electron-phonon transport in NTs. While the results reported here shed light on energy dissipation in NTs in the diffusive regime for which hot electrons interact with optical phonons, further improvement in the sensitivity of the SThM method can potentially allow for the investigation of the transport and energy dissipation mechanisms in SWNTs in the ballistic regime at lower voltage bias.

## ACKNOWLEDGMENT

The authors acknowledge Sergei Plyasunov's contribution during the early stage of this work, and thank Arden Moore for calculating the spreading resistance. This work is supported in part by Department of Energy award DE-FG02-07ER46377, National Science Foundation Thermal Transport Processes Program, and Texas Higher Education Coordinating Board Norman Hackerman Advanced Research Program.

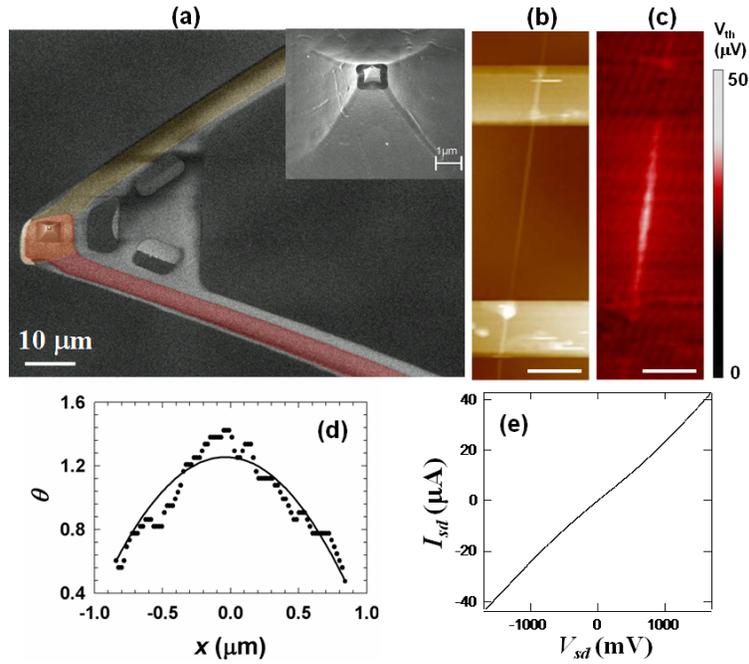

Fig. 1. (a) Scanning electron micrographs (SEMs) of the scanning thermal probe. False colors are added to highlight the Pt (yellow) and Cr (yellow) line (b) AFM height image of a 10 nm diameter MWNT and two metal contacts. Scale bar: 500 nm. (c) DC thermal image at 0.73 V bias and 20.1 μA current. Scale bar: 500 nm. $V_{th}$ is the thermovoltage of the thermal probe. (d) Dimensionless temperature ($\theta$) profile (symbols) along the NT from c. The solid line is the least square fitting using Eq. 2 with the fitting parameters $m = 0.16\ \mu m^{-1}$ and $q^* = 79.2$. (e) Source-drain current ($I_{sd}$) of the MWNT as a function of the source-drain voltage ($V_{sd}$) at zero gate voltage ($V_g$).

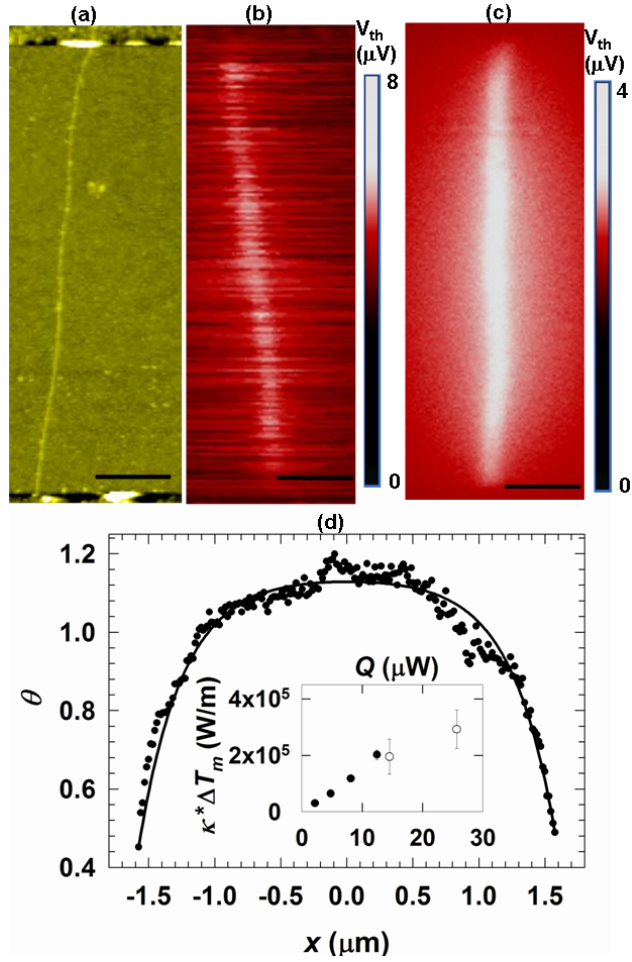

Fig. 2. (a) AFM height image of a 1.8 nm diameter metallic SWNT. (b) DC thermal image with 1.7 V bias and 15.1 µA current. (c) First harmonic AC thermal image with bias of $(0.767 + 0.733 \sin 2\pi ft)$ V and $f = 205$ Hz. Scale bars in a-c: 500 nm. The tip scan directions were slightly different between the three images in a-c. (d) Dimensionless temperature ($\theta$) profile (symbols) along the NT from c. The solid line is the least square fitting using Eq. 2 with the fitting parameters $m = 3.0097$ µm$^{-1}$ and $q^* = 1.1405$. Inset: $\kappa^* \Delta T$ obtained from AC (filled circles) and DC (unfilled circles) thermal imaging results as a function of electrical heating rate ($Q$). The uncertainty of the AC results is smaller than the symbol size.

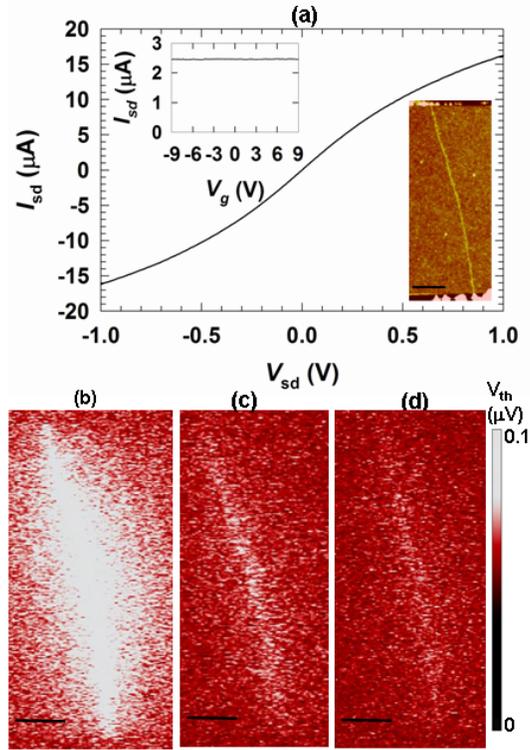

Fig. 3. (a) $I_{sd}$ of a 1.2 nm diameter SWNT sample as a function of $V_{sd}$ at zero $V_g$. The low bias resistance of this sample was 41 kΩ. The upper left inset is $I_{sd}$ as a function of $V_g$ at $V_{sd}$ = 100 mV, which shows that the tube is metallic. The lower right inset is the AFM height image. Also shown are first harmonic AC thermal images at (b) $V_{sd}$ = 0.404+0.123sin$2\pi f t$, (c) $V_{sd}$ = 0.298+0.06sin$2\pi f t$, and (d) $V_{sd}$ = 0.217+0.06sin$2\pi f t$. $f$ = 205 Hz. All scale bars are 500 nm.

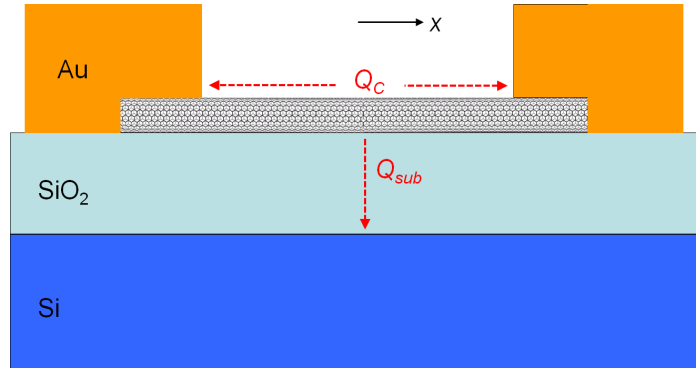

Fig. 4. Schematic illustration of heat flow directions in a current-carrying NT device. $Q_C$ is the heat conduction along the NT to the two electrodes. $Q_{sub}$ is the heat transfer across the NT-substrate interface.